\documentstyle[aps, prbbib]{revtex}
\begin{document}
\draft
\title{Suppression of hole-hole scattering in GaAs/Al$_{0.5}$Ga$_{0.5}$As
heterostructures under uniaxial compression}
\author{V.Kravchenko, N.Minina, A.Savin}
\address{Physics Faculty, Moscow State University, 119899\\
Moscow, Russia }
\author{C.B.Sorensen, O.P.Hansen}
\address{Oersted Laboratory, Niels Bohr institute, Copenhagen }
\author{W.Kraak}
\address{Institute of Physics, Humboldt University, Berlin }
\maketitle

\begin{abstract}
Resistance, magnetoresistance and their temperature dependencies have been
investigated in the 2D hole gas at a [001] p-GaAs/Al$_{0.5}$Ga$_{0.5}$As
heterointerface under [110] uniaxial compression. Analysis performed in the
frame of hole-hole scattering between carriers in the two spin splitted
subbands of the ground heavy hole state indicates, that h-h scattering is
strongly suppressed by uniaxial compression. The decay time $\tau _{01}$ of
the relative momentum reveals 4.5 times increase at a uniaxial compression
of 1.3 kbar.
\end{abstract}


\section{Introduction}

In the middle of the eighties, when a successful growth of perfect
modulation doped p-GaAs/AlGaAs heterostructures initiated an intensive study
of 2D hole systems, a strong positive magnetoresistance of 2D holes confined
in an asymmetric triangular quantum well (QW) was observed in the region of
low magnetic fields \cite{a1,a2}. The lack of inversion symmetry in a QW of
this kind causes lifting of the spin degeneracy of the hole states at k$\neq 
$0, i.e. splitting into two non spindegenerated subbands with different
effective masses, sticking each other at k=0 \cite{a3}. In this connection
the effect of positive magnetoresistance seemed to be associated with
two-band carrier conductivity \cite{a1,a2}, although its strong temperature
dependence remained to be a puzzle. Recently it was found \cite{a4} that
this puzzle can be successfully removed for p-GaAs/AlGaAs heterostructures
by using the model of temperature dependent mutual scattering of the holes
(h-h scattering) in the two non spindegenerate subbands.

In the present paper we show that this h-h scattering mechanism is strongly
suppressed by uniaxial compression. We report on the resistance,
magnetoresistance and their temperature dependencies in the 2D hole gas at a
p-GaAs/Al$_{0.5}$Ga$_{0.5}$As heterointerface in the low and intermediate
magnetic field range under uniaxial compression. Shubnikov-de Haas (SdH)
oscillations and Hall effect were also studied in magnetic fields up to 3.5T
in order to determine the carrier concentrations. We analyze these data in
the frame of a two-band model with h-h scattering as it was done in Ref.\cite
{a4}.

\section{Experimental results}

The samples are processed in the same way and from the same wafer as the
ones reported on in Refs.\cite{a5,a6}, where the emphasis was put on the
range of high magnetic fields, and where uniaxial pressure dependence of the
effective mass m$_1$ as well as the carrier concentrations n$_0$ and n$_1$
in the two subbands ''0'' and ''1'' were obtained from SdH and quantum Hall
effects. The wafer is a modulation doped GaAs/Al$_{0.5}$Ga$_{0.5}$As
heterostructure grown by molecular-beam epitaxy on a [001] semi-isolating
GaAs substrate, and doped with Be in part of the Al$_{0.5}$Ga$_{0.5}$As. The
uniaxial compression is applied along the [110] direction of a Hall bar
mesa, cf. Ref.\cite{a5} for the experimental details.

The total carrier concentration N is determined by classical Hall effect $%
\left( \rho _{xy}=\frac B{Ne}\right) $ in magnetic fields up to 3.5 T, while
the hole concentration n$_1$ in the more light and less populated spin
subband ''1'' is determined by SdH oscillations. The concentration in
subband ''0'' is obtained as n$_0$=N-n$_1$. The pressure dependent values of
N, n$_0$, and n$_1$ correspond well to the data from \cite{a5,a6} and are
used as input parameters in calculations of the 2D holes mobilities and of
the mutual scattering characteristics. Galvanomagnetic characteristics,
taken in low and intermediate magnetic fields $\mu $B$\leq $10 and in the
temperature interval 1.7-4.2K, are represented on Figs.1, 2 and show the
following features:

1. At P=0 the R(B) dependence demonstrates a well pronounced positive
magnetoresistance, that tends to saturation in the region $\mu $B$\simeq $5.
The positive magnetoresistance strongly decreases with uniaxial compression
and almost disappears at p=2.0 kbar (Fig.1a). At P$>$1.3 kbar, where
positive magnetoresistance drastically drops, a negative magnetoresistance
becomes well noticeable in intermediate magnetic fields at B$>$0.5 T
(Fig.1a).

2. In the pressure interval where the positive magnetoresistance is still
well pronounced, it reveals a strong temperature dependence, that
practically disappears in the saturation region (Fig.2).

3.The resistance in zero magnetic field noticeably depends on temperature,
even at T$<$4.2K. Under compression this dependence strongly decreases (Fig.
1 b).

4. In accordance with the previous results\cite{a5,a6}, the electrical
resistivity R of the 2D hole gas in zero magnetic field reveals more than 2
times decrease at a uniaxial compression of P=2.6 kbar, while the total
carrier concentration exhibits about 10\% decrease on the background of the
carriers redistribution between the two spin subbands (Figs. 3a, 3b).

\section{Application of hole-hole scattering model}

The contribution of carrier-carrier scattering to electrical resistance is
possible when two types of carriers with different mobilities make up the
electric current. In an electric field the carriers will acquire different
velocities, and the velocity difference can be degraded by carrier-carrier
scattering, which may be described in terms of mutual friction. By writing
the electric current as a sum of two terms: one proportional to the total
momentum and the other proportional to the relative momentum, Kukkonen and
Maldague \cite{a7} demonstrated how the conservation of momentum (the total
momentum) goes along with the mentioned contribution to the electrical
resistance. In the Drude model we then have two coupled vector equations:

\begin{equation}
\frac{m_0V_0}{\tau _0}=eE+eV_0\times B-\eta n_1\left( V_0-V_1\right) \ 
\label{one}
\end{equation}

\begin{equation}
\frac{m_1V_1}{\tau _1}=eE+eV_1\times B-\eta n_0\left( V_1-V_0\right)
\label{two}
\end{equation}

where the subscripts ''0'' and ''1'' characterize each of the two types of
carriers. Comparing these equations to the corresponding equations in
Kukkonen and Maldague\cite{a7}, it is found that the ''friction coefficient $%
\eta $'' is expressed as:

\begin{equation}
\eta =\frac{m_0m_1}{(n_0m_0+n_1m_1)\tau _{01}}  \label{tree}
\end{equation}

where $\tau _{01}$ is the decay time for the relative momentum.

Solving Eqs. (\ref{one}) and (\ref{two}) for the velocity components and
using the expression for the current density $j=n_0eV_0+n_1eV_1=\sigma E$ we
find the components $\sigma _{xx}$ and $\sigma _{xy}$ of the conductivity
tensor to be given by the same expressions that were found in Ref.\cite{a4}:

\begin{equation}
\sigma _{xx}=\frac{\left[ N\omega \left( Be\right) ^2+\left( \eta N\omega
+\omega _0\omega _1\right) \left( \eta N^2+n_0\omega _1+n_1\omega _0\right)
\right] e^2}{\left( Be\right) ^4+\left[ N^2\eta ^2+2\eta \left( n_0\omega
_1+n_1\omega _0\right) +\omega _0^2+\omega _1^2\right] \left( Be\right)
^2+\left( \eta N\omega +\omega _0\omega _1\right) ^2}  \label{four}
\end{equation}

\begin{equation}
\sigma _{xy}=\frac{\left[ N\left( Be\right) ^2+N^3\eta ^2+2\eta N\left(
n_0\omega _1+n_1\omega _0\right) +n_0\omega _1^2+n_1\omega _0^2\right] Be^3}{%
\left( Be\right) ^4+\left[ N^2\eta ^2+2\eta \left( n_0\omega _1+n_1\omega
_0\right) +\omega _0^2+\omega _1^2\right] \left( Be\right) ^2+\left( \eta
N\omega +\omega _0\omega _1\right) ^2}  \label{five}
\end{equation}

\begin{equation}
\text{where\ }\omega _i=\frac{m_i}{\tau _i}\ \ \text{and \ }\omega =\frac{%
n_0\omega _0+n_1\omega _1}N  \label{six}
\end{equation}

Finally, the diagonal resistivity element is calculated from

\begin{equation}
\rho _{xx}=\frac{\sigma _{xx}}{\sigma _{xx}^2+\sigma _{xy}^2}  \label{seven}
\end{equation}

We have already described the way in which the total carrier concentration N
and the concentrations n$_0$ and n$_1$ in the two subbands were determined
from Hall effect and SdH measurements. The remaining parameters of the model
were evaluated from the expressions for the high-field saturation value of $%
\rho _{xx}$ :

\begin{equation}
\rho _{xx}=\frac \omega {Ne^2}\ \ \text{for\ }\mu B>>1  \label{eight}
\end{equation}

and the zero-field value:

\begin{equation}
\rho _{xx}=\frac{\eta (T)N\omega +\omega _0\omega _1}{\left( \eta
(T)N^2+n_0\omega _1+n_1\omega _0\right) e^2}\ \ \text{for\ }B=0  \label{nine}
\end{equation}

For the high-field saturation value we neglected the slight temperature
dependence, cf. Fig.2, and calculated $\omega $ from Eq.(\ref{eight}).
Afterwards we calculated $\eta $ from Eq.(\ref{nine}) at each of the
experimental temperatures for an array of $\omega _1$-values ($\omega _0$
was eliminated using Eq.(\ref{six})). Each $\omega _1$ -value thus gave the
friction coefficient as function of temperature. We finally determined the
best value of $\omega _1$ as the one which gave the best fit of $\eta $ to
the relation

\begin{equation}
\eta (T)=\alpha T^2  \label{ten}
\end{equation}

which is the expected temperature dependence when Fermi-Dirac statistics is
prevailing; i.e. when kT$<<E_F$ \cite{a8}. In the samples under
investigation $E_F\simeq 6meV$. The resulting parameter values $\omega _0$, $%
\omega _1$ and $\alpha $ are displayed in Figs.3c and d. In Fig.3c we have
replaced $\omega _0$ and $\omega _1$ by the corresponding mobilities.

\section{Results and discussion}

The behavior of the magnetoresistance at different pressures and
temperatures has been calculated from expressions (\ref{four}),(\ref{five})
and (\ref{seven}) with the obtained parameter values $\omega _0$, $\omega _1$
and $\alpha $. It is depicted on Figs.1 and 2 by dotted curves. The maximal
deviation of calculations from the experimental curves does not exceed 10\%
for $\Delta R=R(B)-R(0)$ in the whole interval of magnetic fields, pressures
and temperatures under investigation. Thus we can conclude, that the
experimental temperature dependence of the magnetoresistance at different
uniaxial pressures can be well described by mutual scattering of holes in
the two spin-subbands. The temperature dependence of the resistivity in zero
magnetic field does not follow the h-h scattering model calculations (dotted
curves) at T$>$5K (see insert on Fig.1b). It means that at higher
temperature this model is not suitable because of growing kT and scattering
on acoustic phonons. It should be noted that the calculations were performed
only for the pressure interval up to 1.3 kbar, because the noticeable
negative magnetoresistance (Fig.1), which origin is not clear at present,
introduces an apparent deviation from the two-band model, described by
expressions (\ref{four}),(\ref{five}) and (\ref{seven}). The pressure
dependence of the mobilities $\mu _0$ and $\mu _1\ \left( \mu _i=\frac e{%
\omega _i}\right) \ $in the two spin subbands reveals their increase under
uniaxial compression (Fig.3c), while the value of $\alpha $, that describes
the mutual friction coefficient $\eta =\alpha T^2,$ strongly decreases
(Figs.3d). The decay time $\tau _{01}$ of the relative momentum at different
magnitudes of uniaxial compression may be estimated with the help of Eq. (%
\ref{tree}), using the experimental values for m$_1$ and parabolic
approximation for m$_0$ from Ref.\cite{a6}. The values of $\tau _{01}$, that
characterize the mutual scattering of holes in the two spin subbands, as
well as the relaxation times $\tau _0$ and $\tau _1$ are represented in
Table I for T=4,2K. At zero pressure $\tau _0$, $\tau _1$ and $\tau _{01}$
are of the same order of magnitude, but under uniaxial compression $\tau
_{01}\ $reveals a much more fast increase, indicating strong suppression of
the h-h scattering.

\begin{table}[tbp]
\caption{Spin subband relaxation times $\tau _0$, $\tau _1$\ and relative
momentum decay time $\tau _{01}$\ at T=4.2K for different magnitudes of
uniaxial stress.}
\begin{tabular}{cccc}
Stress, kbar & $\tau _0,ps$ & $\tau _1,ps$ & $\tau _{01},ps$ \\ \hline
0 & 5 & 7 & 11 \\ 
0.65 & 6 & 9 & 18 \\ 
1.0 & 8 & 10 & 30 \\ 
1.3 & 11 & 11 & 47 \\ 
&  &  & 
\end{tabular}
\end{table}

We estimated also the decay time $\tau _{01}$ for a 2D-hole system with
parameters n$_0$, n$_1$, $\omega _0$, $\omega _1$ and $\alpha $ from Ref.%
\cite{a4} using the values of effective masses m$_0$, m$_1$ from Ref.\cite
{a9}, where the total carrier concentration of 2D-holes at similar
GaAs/AlGaAs heterointerface is close to the one from Ref.\cite{a4}. In this
case the magnitude $\tau _{01}$=2ps at T=4.2K is about 6 times less than our
value from Table I. This difference is most probably connected with 2$\div $%
3 times less total concentration (and Fermi energy correspondingly) of
2D-holes in Ref.\cite{a4} than we have in our heterostructures.

The present experimental results are in accordance with the previous
conclusion, namely that the splitting of the two heavy hole spin subbands
decreases under uniaxial compression \cite{a5,a6}. Thus, the decrease of the
positive magnetoresistance under uniaxial compression indicates a decrease
of the difference between carrier properties in the two spin subbands. The
hole-hole scattering, that determines the temperature dependence of the
magnetoresistance at different magnitudes of uniaxial compression, does not
reflect subband splitting directly. However, it is known that h-h (e-e)
scattering in one-type carrier systems does not contribute to resistivity 
\cite{a8}. In accordance with this, in rectangular QW's, where subband
splitting is not expected, positive magnetoresistance is almost negligible 
\cite{a1}.

A further result of the analysis is connected with the increase under
compression of the mobilities in the two spin subbands (Fig.3c). According
to Ref.\cite{a6} the effective mass $m_1$ reveals an increase under uniaxial
compression of 2.2 kbar and is therefore not responsible for the increase of
the mobility in this subband. We are thus led to consider the growth of the
mobilities to be connected with a decrease of the scattering on remote
charged impurities. Such impurities exist in the active layer of Al$_{0.5}$Ga%
$_{0.5}$As according to the modulation doped MBE technique, and may be
supposed to be influenced by uniaxial compression. The presence of deep
states in the energy gap of Al$_{0.5}$Ga$_{0.5}$As doped with Be has been
stated in Refs.\cite{a10,a11}.

\section{Summary}

In summary, we have observed a significant influence of uniaxial compression
along [110] direction on zero field resistivity and magnetoresistance of 2D
holes in an asymmetric [001] triangular QW as well as on the temperature
dependencies of these quantities. The experimental results can be well
described in the frame of the classical two-band model, where the two
splitted subbands of the ground heavy hole state constitute the two bands of
the model, and where temperature dependent mutual scattering between the
holes in these bands is taken into account. The results of our calculations
indicate, that the hole-hole scattering mechanism in the 2D hole system
under investigation is strongly suppressed by uniaxial compression. The
results are in qualitative agreement with our previous findings, that the
subband splitting decreases under [110] compressive strain \cite{a5,a6}.

\section{Acknowledgments}

This work has been supported by grant 97-02-17685 from the Russian
Foundation for Basic Research and by grants 9401081 and 9601512 from the
Danish Research Council. The authors are grateful to Prof. S.Beneslavski for
helpful discussions.

\begin{figure}[tbp]
\caption{ ~Uniaxial compression influence on a) the magnetoresistance at
T=1.7K and b) the temperature dependence of the resistivity of the 2D-hole
gas. Dotted curves are the results of calculations with h-h scattering
mechanism taken into account. }
\end{figure}
\begin{figure}[tbp]
\caption{ ~Temperature dependence of the magnetoresistance at a uniaxial
compression of 0.65 kbar. The results of the theoretical calculations are
represented by the dotted curves on the insert.}
\end{figure}
\begin{figure}[tbp]
\caption{ ~Uniaxial compression dependence of a) total carrier concentration
N and spin subband carrier concentrations n$_0$ and n$_1$, b) resistivity in
zero magnetic field, c) spin subband mobilities and d) friction parameter $%
\alpha $.}
\end{figure}

\end{document}